# Colorectal cancers differ in respect of PARP-1 protein expression

Violetta Sulzyc-Bielicka[1], Pawel Domagala[2], Jolanta Hybiak[2], Anna Majewicz-Broda[2], Krzysztof Safranow[3], Wenancjusz Domagala[2]

[1]Department of Clinical Oncology, Pomeranian Medical University, Szczecin, Poland
[2]Department of Pathology, Pomeranian Medical University, Szczecin, Poland
[3]Department of Biochemistry and Medical Chemistry, Pomeranian Medical University, Szczecin, Poland

Recent findings raise the possibility of PARP inhibitor therapy in colorectal cancers (CRCs). However, the extent of PARP-1 protein expression in clinical specimens of CRC is not known. Using immunohistochemistry we assessed PARP-1 protein expression in tissue microarrays of 151 CRCs and its association with the patient's age, sex, Astler-Coller stage, grade and site of the tumor. High PARP nuclear immunoreactivity was found in 68.2% (103/151) of all cases. In turn, 31.8% (48/151) of tumors showed low PARP expression, including 9 (6%) PARP-1 negative CRCs. There was a significant association of PARP-1 expression with the site of CRC and Astler-Coller stage. A high PARP expression was noted in 79.1% of colon vs. 53.9% of rectal tumors (p = 0.001). The mean PARP-1 score was 1.27 times higher in colon vs. rectal cancers (p = 0.009) and it was higher in stage B2 vs. stage C of CRCs (p = 0.018). In conclusion, the level of PARP-1 protein nuclear expression is associated with the tumor site and heterogeneous across clinical specimens of CRC, with the majority of CRCs expressing a high level but minority – low or no PARP-1 expression. These findings may have a clinical significance because the assessment of PARP-1 expression in tumor samples may improve selection of patients with CRC for PARP inhibitor therapy.

**Key words:** PARP-1, colorectal cancer, PARP inhibitor, colorectal cancer therapy.

## Introduction

Poly(ADP-ribose)polymerase-1 (PARP-1) is a nuclear enzyme activated by DNA breaks and engaged in the repair of DNA single-strand breaks *via* base excision repair pathway (BER) as well as in regulation of transcription and cell cycle [1]. PARP inhibitors disrupt BER rendering this repair pathway inactive what leads to increased levels of persisting single-strand breaks, with DNA double strand breaks (DSB) upon replication as the end result. Tumor cells with defects in homologous recombination, such as *BRCA1*-associated breast cancers, cannot repair DSBs, hence PARP-1 inhibition results in synthetic lethality in *BRCA1* and *BRCA2*-deficient cell lines [2, 3]. Based on these results and preclinical studies that followed, clinical trials of PARP inhibitors in treatment of *BRCA1*-associated and triple negative breast cancers have been initiated [4, 5] although results have not always fulfilled the expectations [5]. One reason may be that a certain percentage of *BRCA1*-associated cancers and triple-negative breast cancers had a low expression of or did not express nuclear PARP-1 protein [6, 7] so lack of the target protein could influence the results.

PARP-1 may play an important role in carcinogenesis of CRC [8, 9]. Based on preclinical studies which showed that PARP inhibitors can enhance the radiosensitivity and synergize with chemotherapy in experimental models of CRCs, several clinical trials have recently been initiated with the use of four PARP inhibitors in patients with CRC (reviewed in [10]). The goal is to elu-cidate the significance of PARP inhibition as a way to sensitize CRC to DNA damaging agents currently used in treatment of CRC. Recently, a novel approach to colon cancer therapy in which 5-fluorodeoxyuridine (FdUrd) is combined with a small molecule PARP inhibitor has been suggested [11]. Such PARP inhibitors sensitized both mismatch repair proficient and deficient colon cancer cell lines to FdUrd because BER turned out to be a critical repair pathway when these cells were exposed to FdUrd. These findings raise a possibility that therapies that combine FdUrd with a PARP inhibitor may act against colon cancer cells despite the fact that they do not have defects in homologous recombination [11]. Thus, in view of incoming possibility of PARP inhibitor therapy in colorectal cancers (CRCs) it is important to reveal the range of expression of PARP protein in clinical samples of this cancer. However, little is known about PARP-1 protein expression in CRC [8, 12], specifically the extent of PARP protein expression in clinical specimens of CRC is not known. On the other hand, it has been pointed out that a low expression or lack of the target protein could influence the interpretation of clinical trials of PARP inhibitors [13].

The purpose of this report is to assess the expression of PARP-1 protein in CRC and to explore its association with several clinicopathological factors such as age and sex of the patient, Astler-Coller stage of the disease, grade and site of the tumor.

## Material and methods

### Patients

The study was based on tumor tissue from 151 unselected, consecutive patients who met the following criteria: 1) had undergone a potentially curative colorectal resection for sporadic CRC (absence of relevant family history at the time of admission to the hospital); 2) had no chemotherapy or radiotherapy prior to the operation; 3) invasive adenocarcinoma Astler-Coller B2 or C without involvement of resection margins was diagnosed by histopathological examination; 4) distant metastases at the time of operation were excluded.

### Tumor pathology and tissue microarray construction

Tumor tissue was fixed in buffered 10% formalin and embedded in paraffin. Sections (4 μm thick) were stained with hematoxylin and eosin for histopathological diagnosis. A pathology review was conducted at the Department of Pathology, Pomeranian Medical University in Szczecin by two pathologists (PD, WD) associated with the study to confirm histological diagnosis of the colorectal cancer type. Representative histological slides were used for tissue microarray construction. Two different regions of tumors in the area of outer invasive margin of cancer with highest mitotic activity were identified and marked on hematoxylin and eosin stained sections. The corresponding areas on the tissue paraffin blocks were cored and transferred to a recipient master block using a Tissue Microarrayer (Beeacher Instruments, Silver Spring, MD, USA). Each core was 0.6 mm wide. The recipient block was cut and sections were transferred to coated slides. One slide was stained with hematoxylin and eosin and a subsequent slide for immunohistochemistry.

### Immunohistochemistry

Slides with TMs were deparaffinized, rehydrated, and endogenous peroxidase activity was blocked. Slides were immersed in pH 6.0 buffer and heat induced antigen retrieval was performed in a water bath at 98°C for 20 min. Monoclonal anti-PARP-1 antibody (F-2, dilution 1 : 500; Santa Cruz Biotechnology, Santa Cruz, CA) was used. Slides were incubated for 30 min and immunostained using the Dako EnVision™+ kit according to the manufacturer's instructions. The reaction was developed with a diaminobenzidine substrate – chromogen solution and slides were counterstained with hematoxylin. Positive staining in stromal lymphocytes served as a built-in positive control. Appropriate positive and negative controls were run. Thus, the immunohistochemical procedure for all TMs from 151 tumors was performed at the same time under identical conditions and with the use of the sensitive Envision™+ visualization system. Detection methods using signal amplification with HRP-labeled polymer (such as Envision™+) have been shown to be more sensitive than methods without such a layer of amplification [14]. CRCs were labeled as proximal or distal in relation to the splenic flexure.

### Scoring

Tumor cores were assessed by 3 observers (JH, AM, WD) who were blinded to clinical and pathological data. In cases of disagreement, the result was reached by consensus. To assess the immunohistochemical expression of PARP-1 we used the multiplicative quickscore method (QS) because it seems to be the most reliable and proved to be useful and reproducible [15]. Both intensity (0-3) and pattern (1-6) scores were assessed. Each intensity score was multiplied by its corresponding pattern score (1 = 0-4% of positive tumor cells; 2 = 5-19%; 3 = 20-39%; 4 = 40-59%; 5 = 60-79%; 6 = 80-100%) to obtain the final QS. Nuclear PARP-1 expression was graded as low (QS: 0-9) or high (QS: 10-18). In order to reach the QS, all tumor cells in the core of the tissue microarray were counted.

### Statistics

Since the distribution of QS values was significantly different from normal distribution (Shapiro-Wilk test), non-parametric tests were used for the analysis.

Association of QS values with categorical variables was analyzed with Mann-Whitney test. The association between the presence of high PARP nuclear immunoreactivity (QS > 9) and other categorical variables was analyzed with Fisher exact test. A multivariate logistic regression model was used to find the independent factors associated with high QS values. P < 0.05 was considered statistically significant. STATISTICA version 10 (StatSoft Inc., Tulsa, OK., USA) was used for the statistical analysis.

## Results

The mean age of patients was 60.8 ± 10.0 years, with a range of 34-83 years, and a median of 62 years. Table I lists other clinicopathological characteristics of 151 tumors and patients.

There was a wide variation in PARP-1 protein expression in the CRCs. Immunohistochemical staining with the PARP antibody revealed strong nuclear reaction in tumor cells (Fig. 1A, C) in the majority of CRCs. The remaining tumors exhibited a moderate or low level of PARP-1 expression, or were PARP-1 negative (Fig. 1B, D). Figure 1 shows representative examples of PARP-1 positive and PARP-1 negative CRCs. The percentage of PARP-1 positive tumor cells differed among CRCs. In some CRCs there was also a variation in intensity of PARP-1 protein expression among tumor cells. The distribution of nuclear PARP quickscores (QS) among 151 CRCs is shown in Fig. 2. PARP-1 immunoexpression, although of lower intensity, could also be found in the nuclei of some lymphocytes and stromal cells (Fig. 1A, C).

High PARP nuclear immunoreactivity (QS > 9) was found in 68.2% (103/151) of all cases. In turn, 31.8% (48/151) of tumors showed low PARP immunoreactivity (QS ≤ 9) including 9 (6%) CRCs with no immunoreactivity.

There was a statistically significant association between nuclear PARP-1 QS and the site of the CRC. A high PARP expression was detected in a larger percentage of colon cancers (68/86; 79.1%) than rectal tumors (35/65; 53.9%) (p = 0.001), i.e. it was 1.5 times more frequent in colon than in rectal cancer. The mean PARP-1 score was 1.27 times higher in colon vs. rectal cancers (p = 0.009) (Table I). There was also a statistically significant association between nuclear PARP-1 QS and Astler-Coller stage (mean PARP-1 QS in B2 vs. C = 13.0 ±5.4 vs. 10.6 ±6.3, p = 0.018). No other statistically significant associations between PARP-1 expression and clinicopathologic characteristics were found.

When QS >9 vs. ≤ 9 was used as the cut-off level, univariate logistic regression analysis has shown that the odds ratio (OR) of QS > 9 was 3.24 (95% CI: 1.58-6.64, p = 0.001) for colon as compared with rectal tumors. A multivariate logistic regression analysis adjusted for age, sex, grade and Astler-Coller stage demonstrated that the CRC site was the only independent parameter significantly associated with PARP-1 expression (OR = =3.72, 95% CI: 1.75-7.90, p = 0.0006).

Table I. Mean PARP QS* in relation to clinicopathological characteristics

| Parameter | N (%) | Mean ± SD | P |
|---|---|---|---|
| sex | | | |
|     females | 80 (53) | 11.40 ±6.00 | 0.434 |
|     males | 71 (47) | 12.04 ±6.11 | |
| histological grade | | | |
|     G1 + G2 | 83 (56) | 11.42 ±5.97 | 0.411 |
|     G3 + muc | 65 (44) | 12.05 ±6.04 | |
| astler-Coller stage | | | |
|     B2 | 68 (45) | 13.02 ±5.43 | 0.018 |
|     C | 83 (55) | 10.61 ±6.34 | |
| site | | | |
|     colon | 86 (57) | 12.90 ±5.49 | 0.009 |
|     rectum | 65 (43) | 10.12 ±6.42 | |
| site** | | | |
|     distal | 119 (79) | 11.52 ±5.96 | 0.3478 |
|     proximal | 32 (21) | 12.34 ±6.40 | |

*quickscore, **with regard to splenic flexure

## Discussion

The introduction of PARP inhibitors has become an exciting novel approach to treatment of human cancers. However, despite significant progress in understanding the role of PARP-1 in DNA repair and other biological processes, the level and extent of PARP-1 protein expression in major tumor types, including CRC has remained largely unknown and is currently intensively investigated e.g. in breast cancer [7], ovarian cancer [16], hepatocellular cancer [17], pancreatic cancer [18], melanoma [19], and glioblastoma [20]. Specifically it is not clear whether PARP-1 overexpression is a characteristic feature of particular tumors, their subtypes or subpopulations of tumor cells within a given tumor. This knowledge seems to be necessary to properly assess the results of clinical trials involving targeted therapy with PARP-1 inhibitors [13]. However, little is known about the range of PARP-1 protein expression in clinical specimens of CRC. Positive immunohistochemical expression of PARP-1 was found in 97.7% (42/43) of CRCs [21], 82.1% (23/28) of CRCs [12], and 58.3% (35/60) of colorectal adenomas and pT1 CRCs (however, the exact percentage of CRCs that expressed PARP-1 protein was not given in this report) [8]. An association between PARP-1 protein expression and clinicopathological char-acteristics such as age and sex of the patient, Astler-Coller stage of the disease, grade and site of the tumor has not been studied to date.

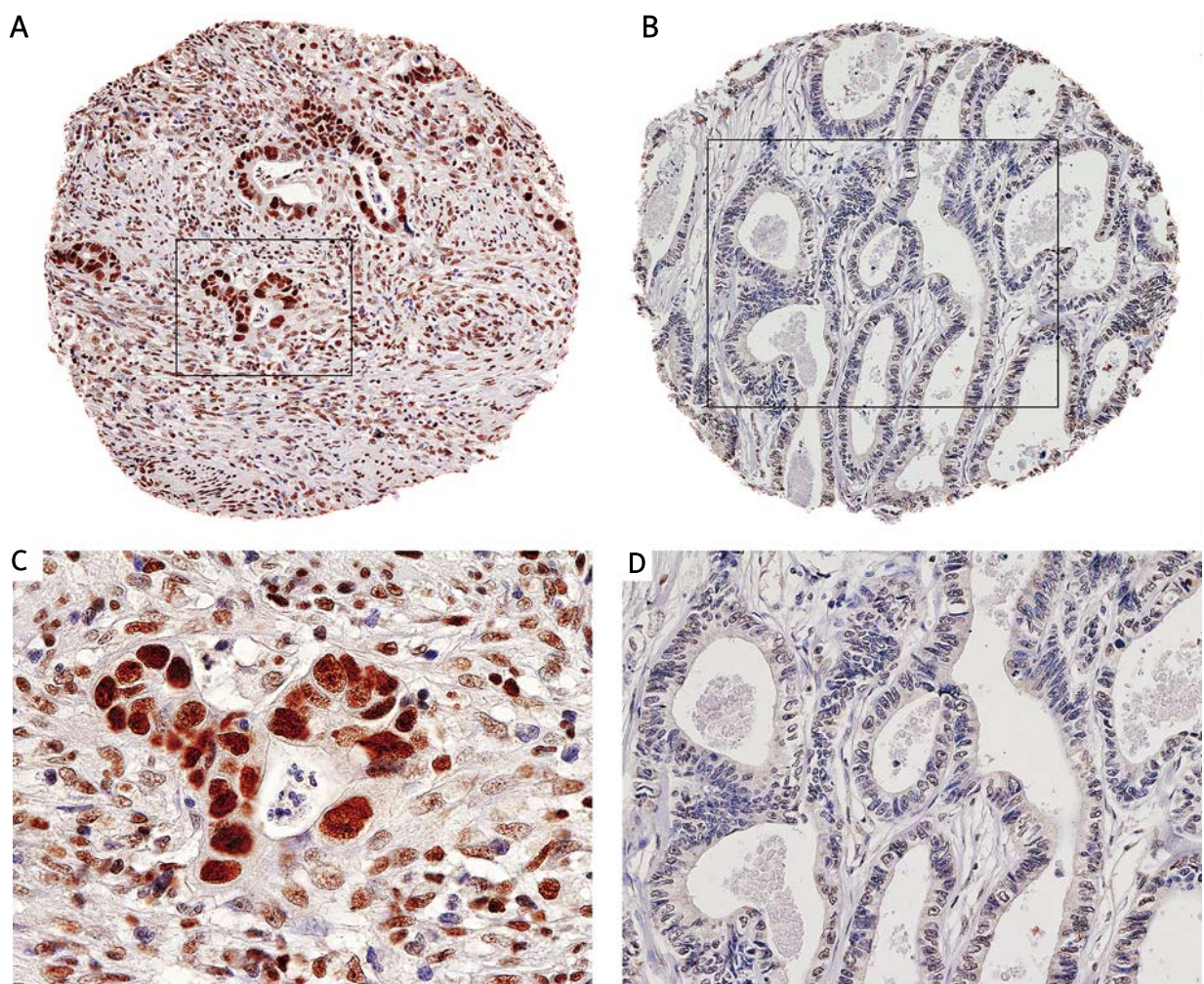

**Fig. 1.** PARP-1 expression in colorectal cancers (immunohistochemical reaction with PARP-1 monoclonal antibody). A and B. Two cores from a tissue microarray, one (A) with PARP-1 protein expression in nuclei of tumor cells, the other (B) negative for PARP-1 protein (original magnification 100×). C and D. High magnification (400×) of boxed area from A (strong nuclear PARP-1 expression) and B (PARP-1 negative tumor cells)

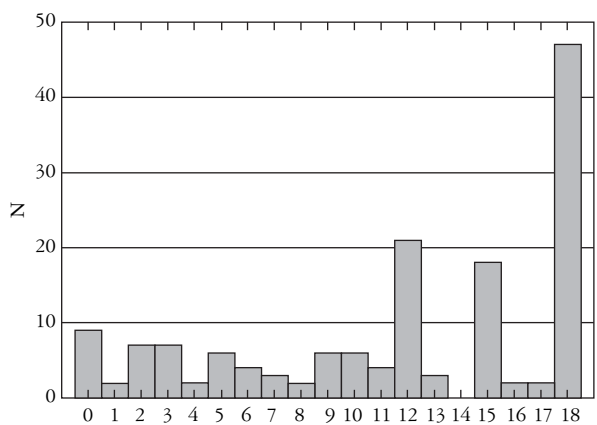

**Fig. 2.** Distribution of quickscores (QS) of nuclear PARP-1 expression in 151 colorectal carcinomas

In this report we evaluated PARP-1 expression in 151 CRCs in order to provide detailed information on PARP-1 protein expression in this cancer and at the same time to broaden our still incomplete knowledge about PARP expression in clinical specimens of human primary cancers. We found that:
- the majority of CRCs exhibited a high expression of nuclear PARP-1 protein;
- CRCs constitute a heterogeneous group of tumors with respect to PARP-1 expression, with some tumors expressing high and other cancers – lower levels of this protein;
- approximately 6% of CRCs did not exhibit PARP-1 protein expression, and altogether, 32% of all tumors exhibited low PARP-1 protein expression. Low or no PARP-1 protein expression may have important therapeutic implications because such tumors may be refractory to treatment with PARP inhibitors (due to lack of the target protein) and therefore they may require different therapeutic approaches. Clinical be-havior including response to treatment of PARP-1 protein negative CRCs is currently unknown and requires further studies. In this context it is of interest that $PARP-1^{-/-}$ cells display enhanced genomic instability including chromosome aberrations, both in the absence of and after DNA damage [22, 23];

- nuclear PARP-1 protein expression was significantly associated with the tumor site (colon vs. rectum) and Astler-Coller stage (B2 vs. C).

We also noticed that expression of PARP-1 was significantly increased in cancer cells as compared with normal stromal cells. A similar phenomenon has been reported in a small group of patients with CRCs [21]. A slightly elevated level of PARP-1 mRNA was also found in a group of 26 colon cancers as compared to normal colon mucosa [24].

Loss of PARP-1 activity following treatment with PARP inhibitors may induce synthetic lethality of tumor cells whose homologous recombination-dependent DNA double strand breaks repair is impaired because of mutations in *BRCA1* or *BRCA2* [2]. For example, deficiency in DNA repair through homologous recombination, which is a characteristic feature of *BRCA1/2*-associated tumors, has been exploited in treatment of breast cancers with PARP-inhibitors. However, it seems that germline mutations are not necessary to obtain the therapeutic effect because *BRCA1*-non-related tumors may also constitute a potential target for some PARP inhibitors [25]. Recently it has been suggested that CRCs displaying microsatellite instability and deficient in double strand break repair due to mutation in *MRE11* gene (which is involved in DNA repair through homologous recombination) can show a higher sensitivity to PARP-1 inhibition, i.e. they also can exhibit a synthetic lethality phenomenon [26]. Therefore, PARP positive CRCs might also be sensitive to such treatment. Another recent report suggests that small molecule PARP inhibitors sensitize colon cancer cell lines to FdUrd irrespective of their mismatch repair status [11]. This raises a possibility that therapies that combine FdUrd with a PARP inhibitor may act against colon cancer cells despite the fact that they do not have defects in homologous recombination [11]. In view of those results the information on PARP protein expression in CRCs seems to be of utmost importance for designing clinical trials with the use of PARP inhibitors and proper interpretation of results.

In conclusion, we showed here for the first time that the level of PARP-1 protein nuclear expression is associated with the site of CRC and is heterogeneous across clinical specimens of CRC. We also showed that the majority of CRCs expressed a high level of nuclear PARP-1 protein but minority of tumors exhibited a low level of PARP-1 protein expression. The findings seem to indicate that a subset of patients with low or no PARP-1 protein expression may not respond to PARP inhibitor therapy or may benefit very little. Therefore, we suggest that future clinical trials involving PARP-1 inhibitors in CRC should take into account PARP protein expression in tumor cells. Clearly, further research is needed on the association of PARP-1 protein expression with clinical outcome of patients with CRCs treated with PARP inhibitors.

*This study was supported by Pomeranian Medical University Research Program grant no. WL-125-01/S/11.*

### Address for correspondence

Prof. **Wenancjusz Domagala** MD
Department of Pathology
Pomeranian Medical University
ul. Unii Lubelskiej 1
71-252 Szczecin, Poland
tel./fax +48 91 487 00 32
e-mail: wenek@pum.edu.pl